\documentclass[11pt]{article}

\usepackage{graphicx}
\usepackage{latexsym}
\usepackage{amstext}
\usepackage{amsxtra}

\begin{document}

\title{Efficient magneto-optical trapping of a metastable helium gas}

\author{
F. Pereira Dos Santos, F. Perales$^{a}$, J. L\'eonard, A. Sinatra, Junmin Wang$^{b}$,\\
F. S. Pavone$^{c}$, E. Rasel$^{d}$, C.S. Unnikrishnan$^{e}$, M. Leduc\\
Laboratoire Kastler Brossel$^*$, D\'epartement de Physique, Ecole Normale Sup\'erieure, \\
24 rue Lhomond, 75231 Paris Cedex 05, France}                    

\date{}

\maketitle

\abstract{
This article presents a new experiment aiming at BEC of metastable helium atoms. It describes the design of a high flux discharge source of atoms and a robust laser system using a DBR diode coupled with a high power Yb doped fiber amplifier for manipulating the beam of metastable atoms. The atoms are trapped in a small quartz cell in an extreme high vacuum. The trapping design uses an additional laser (repumper) and allows the capture of a large number of metastable helium atoms (approximately $10^9$) in a geometry favorable for loading a tight magnetostatic trap.
\

PACS{
      {32.80.Pj}{Optical cooling of atoms;trapping}   \and
      {}{}
     } 
}

\maketitle

\section{Introduction}
\label{sec:intro}

The main goals of this experiment is first to produce a gas of $^4$He atoms in the metastable $2^{3}S_1$ state with a density as large as possible, and then to bring down the temperature of the gas to ultralow temperatures. This experiment replaces an earlier experiment based on the VSCPT cooling method \cite{Sauba99} which, despite the achievement of ultra low temperatures, could not reach high phase space densities, due to the small number of atoms being cooled. 
Let us mention that several other groups are involved in experiments with similar goals dealing with trapped ultracold metastable helium atoms \cite{Rooi97,Tol99,Novak00}. The present experiment, as well as those of references \cite{Tol99,Novak00}, aim at achieving Bose-Einstein condensation (BEC) by combination of laser cooling, magnetic trapping and subsequent evaporative cooling, following the route successfully taken for BEC of alkali atoms (Rb,Na,Li) since 1995 \cite{Anderson95,Bradley957,Davis95} and atomic hydrogen in 1998 \cite{Fried98}. 
A metastable helium condensate would be the first one with atoms in an excited state of high internal energy (19.8 eV) and long lifetime (approximately 8000 s). It should be interesting to compare the properties of such a gaseous dilute helium condensate to those of superfluid liquid helium, dominated by interactions between particles. Helium BEC should be capable of forming a helium atom laser as it was the case for alkali BEC. There are many applications of such coherent matter waves. For instance a metastable atom laser \cite{Mom97} could be a valuable source for lithography \cite{Bald98}. Another application for helium atoms is metrological \cite{Minardi99}, since its energy levels can be calculated with a high degree of accuracy. Let us finally mention an interesting property of these metastable helium atoms. They can transfer their high internal energy when they collide with surfaces or molecules. This property can be used for highly efficient detection, almost "one by one", with good spatial and temporal resolution \cite{Mlynek97}, using microchannel plates for example.

It thus appears that the metastable helium atom displays appealing features as a candidate for BEC. According to theoretical predictions \cite{Shlyap96}, the cross section for elastic collisions between cold metastable atoms should be large, ensuring rapid thermalisation for efficient evaporative cooling in the magnetic trap. Furthermore, Penning collisions which are the main source of inelastic losses in a magneto-optical trap, are expected to have a rate slow enough to allow the formation of BEC in a magnetostatic trap. However, reaching BEC with metastable helium remains uncertain. The predicted value of approximately 10 nm for the scattering length could be inaccurate as it is very sensitive to the details of long range elastic potential between atoms, which is not known to a high accuracy. It should also be mentioned that the rates of the inelastic collisions, which are likely to heat up and empty the trap at low temperatures \cite{Shlyap96,Vent99,Beij00}, have not yet been measured.
From an experimental point of view, one first needs to trap a dense cloud of these atoms in an ultra high vacuum. A second step is to construct a strongly confining magnetostatic trap. Thus, we choose to trap the atomic cloud in a quartz cell of small dimension while having the confining magnets external to the cell and close to the vacuum chamber.
Section \ref{sec:source} of this article gives details of the discharge source and of the optimal parameters chosen to achieve a high flux of metastable helium atoms. Section \ref{sec:laser} shows the laser system consisting in a DBR diode laser coupled with a high power fiber amplifier. Section \ref{sec:colldefl} describes the laser techniques used to increase the brightness of the beam of atoms, to deflect it from the beam of ground state atoms and to slow it down in a spatially varying magnetic field. Section \ref{sec:trap} demonstrates the advantages of our particular trapping scheme.

\section{The source of metastable atoms}
\label{sec:source}
\subsection{Principle}
The development of an intense and slow beam of metastable helium atoms requires to solve several problems. First, helium atoms have to be efficiently excited to the metastable state. Secondly, the beam has to be cooled to a low enough temperature to avoid difficulties with the subsequent deceleration, knowing that the small mass of the atom results in a high velocity at room temperature. In a previous experiment \cite{Sauba99}, metastable atoms were produced at a moderate rate by electron bombardment and they were cooled down to liquid helium temperature. In the present setup, a different strategy is used: metastable atoms are produced in a gas discharge and cooled to liquid nitrogen temperature, reaching mean velocities of approximately 1000 m/s.
It is known that the most efficient way to produce high rates of metastable helium atoms is to start with a pulsed or continuous gas discharge, where atoms are excited to upper states by electronic collisions and then decay to the long-lived metastable state. A fraction between $10^{-6}$ and $10^{-4}$ of the ground-state atomic flux can be produced in the metastable state with an intense discharge. However, the heat generation in the discharge makes it difficult to obtain both intense and cold atomic beam of metastable helium. Several attempts \cite{Fahey80,Kawa93,Rooi96} have been made to solve this problem. The source developed at ENS results from a design which combines several advantages of the sources developed by the other groups. The setup is compact and robust and gives a reliable large flux of atoms, a significant number of which have velocities below 1000 m/s.\\
All discharge sources for metastable helium consist basically of a gas reservoir filled with helium gas to a pressure of a few tens of mbar. The discharge occurs between the cathode inside the reservoir through the outlet channel to the anode, placed on the high vacuum side, or directly to the skimmer. The design of Fahey et al. \cite{Fahey80} achieves low gas temperatures by cooling only the nozzle with liquid nitrogen. The source described by Kawanaka \cite{Kawa93} makes use of an elaborate scheme to cool down all of the source and to remove the hot gas by a roughing pump. Similar results have been obtained regarding fluxes reaching up to $10^{14}$ atoms per second and steradian \cite{Fahey80} and \cite{Kawa93}. Velocities are found to be slightly above 1000 m/s in \cite{Fahey80} and slightly below 1000 m/s in \cite{Kawa93}.
For the design presented here and shown in fig. \ref{fig:source}, the complete source, including the discharge electrodes, is cooled in a simple and efficient way. No additional effort such as removing the hot gas is required. As a result the entire source is very compact. A careful design of the shape of the electrodes and of the gas outlet ensures that the discharge is partially burning into the high vacuum (see fig. \ref{fig:source}) rather than inside the reservoir. This provides a high flux of metastable atoms, as they do not hit walls at the place where they are produced. The design is such that a reliable and stable operation mode has been achieved for several weeks of continuous operation.
\subsection{The source design}
\begin{figure}[htb]
\begin{center}
\includegraphics[height=4.5cm,width=9cm]{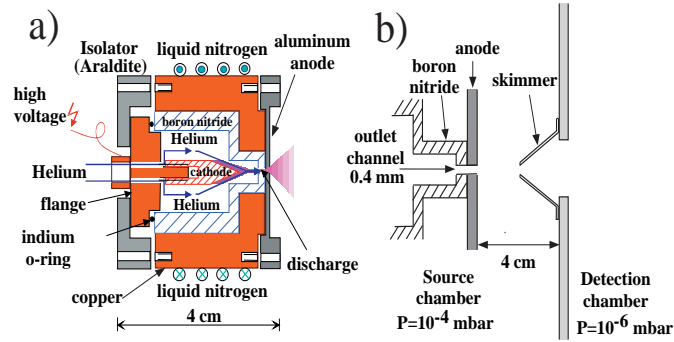}
\end{center}
\caption{\footnotesize a) Compact discharge source of metastable helium atoms. The discharge burns in the outlet channel and outside the anode. The metallic anode is cooled to liquid nitrogen temperature. b) Zoom of the outlet channel region. The source and detection chambers are separated by the skimmer.}
\label{fig:source}
\end{figure}
The source shown in fig. \ref{fig:source} consists of a cylindrical gas reservoir entirely made of boron nitride, a material which has a sufficiently high heat conductivity as well as a high electrical and chemical resistance. This material was already used by Fahey et al. for the nozzle of their discharge source \cite{Fahey80}. The external part of the gas reservoir is covered with vacuum grease for better thermal contact with the copper cylinder cooled by liquid nitrogen, into which it is pressed. A copper flange, held by an electrically isolating support of araldite and tightened by an indium o-ring, closes the reservoir on the backside. The flange serves as the inlet of the helium gas, as well as for the mount of the cathode. The cathode is a stainless steel tip adjustable in position during mounting. The distance between cathode and anode is typically between 2 and 3 mm but it is not found critical. 

The particular shape of both electrodes allows them to be centered with respect to each other by mounting them on the reservoir. Particular care was taken to design the gas outlet, a 2 mm long channel of 0.4 mm in diameter drawn in the boron nitride reservoir and located in front of the cathode. The outlet channel diameter is chosen in such a way that a stable discharge and a high flux of metastable atoms are reached, even at low operating discharge currents and for gas loads adapted to the speed of the pumps. It is directly followed by the anode, a 1 mm thick aluminum disc with a hole of 0.4 mm in diameter. A particular feature of the present design is that the anode is cooled to liquid nitrogen temperature. For efficient cooling it is tightly fixed on the copper container. Crucial for reliable operation of the discharge is the cleanliness of the cathode and anode. During operation both parts suffer from impurities of the gas or from the oil vapor of the diffusion pumps. Our design allows easy dismantling of the source to clean the electrodes.\\
The material, the size, the depth and the alignment are crucial for an efficient excitation of helium atoms. Best performances were obtained with an aluminum anode with a hole of diameter at least as large as that of the outlet channel in the boron nitride reservoir. This ensures that the electrical field lines reach into the source chamber (see fig. \ref{fig:source} b), so that the discharge extends also into the vacuum at the source exit. The cylindrical design and the length (30 mm) of the reservoir have been chosen such that no parasitic discharge can occur competing with the discharge at the source exit when operated in a pressure range of 10 torrs. The source reservoir is filled with gas via a plastic tube, which isolates the source from the vacuum chamber. A throttle valve in front of the gas inlet keeps the pressure high in the gas tube and suppresses parasitic discharge in the tube. To avoid contamination of the source, the helium gas is filtered with charcoal.

\subsection{Measurement of the atomic flux and velocity}
\begin{figure}[htb]
\begin{center}
\includegraphics[height=3.3cm,width=9cm]{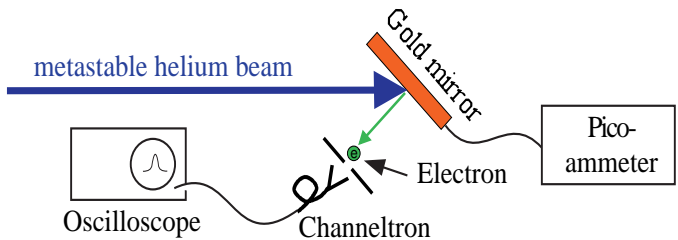}
\end{center}
\caption{\footnotesize In-situ detection of the atomic beam}
\label{fig:detector}
\end{figure}

The source has been tested in a vacuum apparatus consisting of two vacuum chambers (see fig. \ref{fig:source}), one for the source and one for the beam diagnostic. Both chambers are separated by a skimmer (1 mm diameter) and evacuated by diffusion pumps (pumping speed of approximately 800 l/s) equipped with liquid nitrogen traps. During operation the pressure in the source chamber rises to approximately $10^{-4}$ mbar, in the detection chamber to approximately $10^{-6}$ mbar. 
To characterize the atomic beam an in-situ detector was constructed, consisting of a gold mirror and a channeltron (fig. \ref{fig:detector}). Upon collision with the mirror surface, metastable atoms decay down to the ground state and release one electron out of the surface, with a high efficiency \cite{Dunning71}. Assuming that each metastable atom hitting the mirror releases an electron, one gets a lower limit for the atomic flux by measuring the current with a picoammeter. 
The metastable beam can also be pulsed with a mechanical chopper, in order to perform time of flight measurements. In this case, the current on the gold mirror is too weak to be detected. The released electrons need to be accelerated towards a channeltron which detects the amplified current. The pulsed signal of the channeltron is then sent to an oscilloscope to record the time of flight distribution. To separate the metastable triplets $2^{3}S_1$ from other species produced by the discharge (UV light, ions, metastable singlet $2^{1}S_0$), the beam is collimated by diaphragms, and deflected by a laser beam tuned to the $2^{3}S_1 \rightarrow 2^{3}P_2$ transition (see section 3). From the time of flight measurements with and without deflection, it has been observed that the source essentially produces metastable atoms which are in the triplet state. We assume that the singlet state atoms are quenched by the radiation emitted by the discharge.

\subsection{Choice of discharge current and pressure}
\begin{figure}[htb]
\begin{center}
\includegraphics[height=6.3cm,width=9cm]{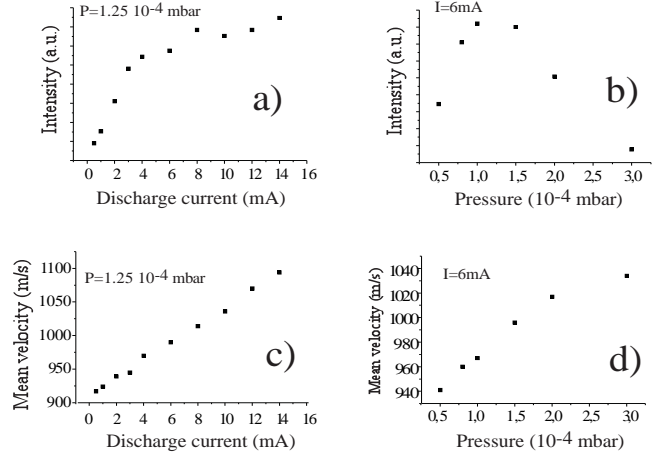}
\end{center}
\caption{\footnotesize Curves a) and b) show the atomic flux, in arbitrary units, inferred from the maximum of the time of flight distribution as a function of the discharge current and the pressure in the vacuum chamber. Curves c) and d) display the corresponding mean velocity as a function of the discharge current and the pressure in the vacuum chamber. Curves a) and c) are taken at a pressure of $1.25\times10^{-4}$ mbar, curves b) and d) for a current of 6 mA. Note : the pressure in the reservoir is proportional to the pressure in the vacuum chamber (see figure 1).}
\label{fig:carac}
\end{figure}

The efficiency of the source depends on a large variety of parameters: the gas pressure and temperature, the discharge current, the purity of helium and the geometry of the discharge. All these parameters have to be carefully optimized to obtain the highest maximum flux with moderate heating. The discharge current and the pressure inside of the gas reservoir were separately varied as shown in fig. \ref{fig:carac}.
For a given pressure, the production rate of metastable atoms increases linearly with the current up to 4 mA (fig. \ref{fig:carac} a). For higher currents the rate starts to saturate. With increasing current, the gas temperature rises locally in the discharge region due to resistive heating resulting in a higher atomic velocity (see fig. \ref{fig:carac} c). A current of approximately 6 mA was chosen as a compromise between high flux and low velocities. In figure 3 b), there are two regimes for the pressure. At lower pressures, an increase of pressure inside the reservoir results in an increase of the metastable helium flux. At higher pressures, an increase of pressure inside the reservoir leads to a decrease of flux due to quenching of metastable helium atoms by collisions between metastable atoms in the nozzle region and by collisions with the background gas. Eventually, if the pressure is increased passed the background pressure of $3\times10^{-4}$ mbar, the metastable helium beam can be completely quenched. For the vacuum setup described above, the optimum pressure was achieved at approximately $10^{-4}$ mbar.
After optimization of all parameters of the atomic source, fluxes of triplet metastable atoms of the order of $2\times10^{14}$ atoms/sec/steradian were found, with a mean velocity of 1000 m/s. Using this highly compact source, the measured values compare well with the ones obtained in other experiments \cite{Fahey80,Kawa93,Rooi96}.

\section{The laser system}
\label{sec:laser}
Earlier experiments on laser cooling and trapping of helium at 1083 nm (transition $2^3S_1 \rightarrow 2^3P_2$) were performed using a LNA ring laser, pumped by an argon ion laser \cite{Vanst91}. In this experiment, we use an optical amplifier based on Ytterbium doped fiber (IRE-POLUS) and seeded by a diode laser at 1083 nm. This laser source is especially efficient to manipulate metastable helium atoms. Historically, the first Ytterbium fiber amplifier was developed and characterized in a single stage low amplification configuration \cite{Pasc96}. Later, a double core prototype of this MOPFA system (Master Oscillation Power Fiber Amplifier) was built by S.V. Chernikov \cite{Chern97}.
\begin{figure}[htb]
\begin{center}
\includegraphics[height=2.4cm,width=9cm]{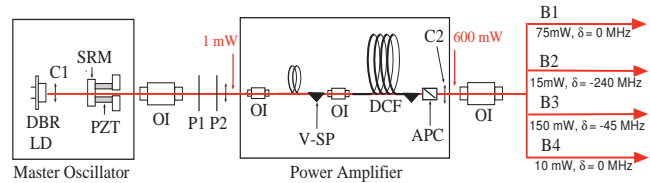}
\end{center}
\caption{\footnotesize Laser setup. A master oscillator at 1.083 $\mu$m (DBR Laser Diode) injects an optical Yb doped fiber power amplifier. C1 : collimator, PZT : piezo-transducer, SRM : semi-reflecting mirror, OI : optical isolator, P1 and P2 : $\lambda/2$ and $\lambda/4$ plates, V-SP : V-groove side-pumping by diode arrays, DCF : double clad fiber, APC : angle polished connector, C2 : collimator. The box representing the power amplifier, commercially available from IRE-POLUS, has two fiber connections for input and output. The output beam is split up in four independent beams : B1 represents the collimation-deflection beam, B2 the slowing beam, B3 the MOT beams and B4 the probe beam.}
\label{fig:lassetup}
\end{figure}
The laser system used in the present experiment is shown in figure \ref{fig:lassetup}. The seed laser is a single mode DBR laser diode (SDL-6702-H1) emitting at 1083nm and delivering a maximum power of 50mW. The line width of the laser diode is reduced from 3 MHz to 250 kHz by an external cavity using a semireflecting mirror of transmission 80 \%, as has already been observed \cite{Pavone95}. The laser diode is coupled with the fiber amplifier using bulk optics. Two optical isolators providing a total isolation of 60 dB prevent optical feedback in the DBR diode. In addition, a set of two birefringent plates ($\lambda$/2 and $\lambda$/4) of adjustable orientation compensates for the birefringence of the amplifier, which slightly varies with temperature changes and mechanical stress. This adjustment provides the proper linear polarization at the output of the amplifier. An additional 30 dB optical isolator is placed at its output because the amplifier is sensitive to feedback from the experiment.
The Yb doped fiber amplifier consists of two amplification stages, both pumped by diode arrays operating at 970 nm (V-groove side pumping). The second amplifying stage (called booster) is designed for 600 mW saturated output power. It consists of a double clad fiber (DCF) with an optimized bidirectional side pumping. The angle polished connector (APC) at the output end prevents the amplifier from oscillating. An input power of 1mW is sufficient to saturate the amplifier and achieve performances independent of the input level. The amplifier provides a collimated beam in a TEM00 mode of 0.4 mm waist.\\
A preliminary study of the frequency noise of the laser source was performed using an autocorrelation setup and heterodyne detection. It was found that the fiber amplifier did not cause additional noise to the frequency spectrum of the injection diode when the diode is frequency narrowed in an external cavity.\\
The laser diode is locked -240 MHz away from resonance by saturated absorption in a low pressure discharge cell. The fiber-laser light is split up into four independent beams (see fig. \ref{fig:lassetup}): the first one is used to collimate and deflect the atomic beam, the second one to slow the beam down, the third one to trap the atoms, and the last one to probe the trap (see sections \ref{sec:colldefl} and \ref{sec:trap}). Required frequency for each arm is set by acousto-optical modulators used in a double-pass configuration. 
 
\section{Laser manipulation: collimation, deflection and deceleration}
\label{sec:colldefl}
\begin{figure*}[htb]
\begin{center}
\includegraphics[height=5cm,width=15cm]{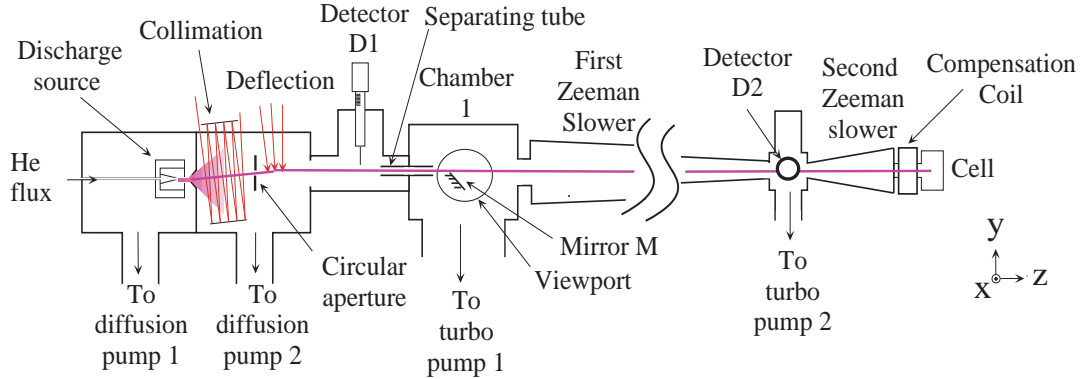}
\end{center}
\caption{\footnotesize Experimental setup. The metastable helium beam is collimated, deflected, decelerated and trapped at the end of the setup in a small dimension quartz cell.}
\label{fig:expsetup}
\end{figure*}

The need for extremely high vacuum in the present experiment requires that the intense ground state helium beam is prevented from reaching the cell. The metastable beam, initially merged with the ground-state beam, has then to be spatially separated and directed towards a different axis than the nozzle-skimmer axis. Radiation pressure forces are used for this purpose \cite{Rasel99}. For the collimation and the deflection, a power of 75 mW of laser light is used (beam B1 in the figure \ref{fig:lassetup}). This power is evenly split in three : the first one for vertical collimation, the second one for horizontal collimation and the last one for deflection. 
The effusive beam coming out from the source is highly divergent (0.1 rd) and has a uniform spatial intensity profile. Collimation is thus performed slightly off axis, $1^{\circ}$ upwards with respect to the horizontal axis. Two apertures (a diaphragm and a tube) placed off axis selectively blocks the ground state beam while allowing the metastable state beam to be deflected by the laser in order to pass through (see fig. \ref{fig:expsetup}). The circular aperture ($\O=5$ mm) and the separating tube ($\O=1$ cm, length 10 cm) are 1.2 m apart from each other. They define the new axis of the experiment starting 5 mm above the nozzle-skimmer one all the way down towards the cell. The separating tube provides differential pumping in the chamber connected to the main slowing magnet and pumped by a turbo molecular pump (1000 l/s). We use a vertically movable Faraday cup ($\O=7$ mm) located 1.15 m downstream from the nozzle to monitor the intensity of the metastable helium beam (detector D1 in figure \ref{fig:expsetup}). For the collimation of the atomic beam in the two transverse directions, we use the so-called "zig-zag" configuration of the laser beam \cite{Rasel99}. It uses a resonant beam ($\O=8$ mm) reflecting between two mirrors ($3\times15$ cm) sligthly tilted from being parallel to cross the atomic beam about 10 times. The capture range of the transverse velocity is approximately 20 m/s. The increase in the metastable flux is measured on a picoammeter (Keithley) connected to the Faraday cup 1. Although "white light" can be used to achieve collimation \cite{Rasel99}, we did not use it to prevent the broadening of the laser-source linewidth which has multiple uses in the experiment. To deflect the collimated beam, we used a curved-wavefront laser beam in a ribbon shape at resonance. The optimised radius of curvature is of 5 m. The deflection keeps the beam collimated with nearly a 100$\%$ efficiency.

The second Faraday cup D2 ($\O=8$ mm) (see fig. \ref{fig:expsetup}), located 2.4 m away form the tube entry, is used to optimize the flux of the collimated-deflected beam. Typical currents measured on D2 are of 20 nA in comparison with 1 nA measured on D1 when the metastable beam is neither collimated nor deflected. This corresponds to a collimated flux of approximately $2\times10^{11}$ atoms/s.
To characterize the velocity of the pure triplet metastable beam, we performed a time of flight measurement. We used a resonant laser beam ($\O=1$ cm) with a light chopper, crossing at right angle with the collimated-deflected metastable beam, and a channeltron mounted besides the Faraday cup (D2). The light beam acting as an atom pusher is hidden for a short time period ($50\mu$s) every 10 ms. We recorded the time of flight spectrum and found a peak velocity of 930 m/s and a relative spread (FWHM) of approximately $30\%$, which is significantly lower than the uncollimated beam : this shows that the collimation-deflection process acts more efficiently on slow atoms because their interaction time with the laser beams is longer.
The metastable helium beam is decelerated by the Zeeman tuning technique \cite{Phil82}. For this purpose, a laser beam with 15 mW of total power is increased to a diameter of approximately 2 cm, with a right circular polarisation and a detuning from resonance of $\delta_{slo}=-240$ MHz. The laser beam enters the cell and propagates anti-parallel to the atom beam. It is resonant with atoms having longitudinal velocity of 1000 m/s at the entrance of the first Zeeman slower. This first Zeeman slower has a length of 2 m, an inner diameter of 2.2 cm and a field of 540 G. At the end of this first Zeeman slower, the atomic velocity is approximately 240 m/s. The second Zeeman slower is approximately 15 cm long with an enclosed tube of 40 mm inner diameter and creates a field going from 0 to -140 G. The atoms are slowed down to a final velocity of 40 m/s as they exit the second Zeeman slower. A compensating coil minimizes the magnetic field leakage from the second Zeeman slower into the cell region.\\ 
Control of the successive decelerations was done by a Doppler sensitive absorption spectroscopy method, using a laser-probe beam crossing the cell with an angle of approximately $20^{\circ}$. Time of flight measurements of the unslowed beam was used to calibrate the Doppler detuning with respect to the atom velocity. Velocity measurements are in good agreement with simulations of the slowing process.

\section{The trapping scheme}
\label{sec:trap}
\subsection{The laser beams geometry}
\begin{figure}[htb]
\begin{center}
\includegraphics[height=4.5cm,width=9cm]{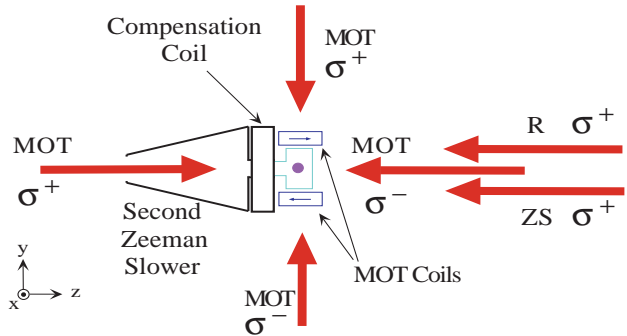}
\end{center}
\caption{\footnotesize MOT setup. The MOT beams are perpendicular to the surfaces of the quartz cell. The trapping scheme requires an extra laser, the repumper (R), which is superimposed to the Zeeman slowing (ZS) and MOT beams along the axis of the atomic beam.}
\label{fig:mot}
\end{figure}
In order to optimize the number of trapped atoms in the MOT, we use a far detuned ($\delta_{mot}/2\pi=-45$ MHz) and high intensity laser (total intensity $I=50$ mW/cm$^2$) (beam B3 in fig. \ref{fig:lassetup}). This laser detuning minimizes inelastic Penning collisions between atoms in the $2^3S_1$ metastable state and atoms in the $2^3P_2$ excited state \cite{Tol99,Kuma99,Brow00,Per00}. Our scheme aims at trapping the gas at the center of a quartz cell of high quality commercially available from Hellma (5cm$\times$5cm$\times$4cm). We use large diameter laser beams ($\O=2$ cm) in order to capture a large number of atoms. The MOT is as close as possible to the slowing magnet end to allow a higher loading rate. The MOT beams are 6 independent laser beams crossing the cell perpendicular to its faces. The two MOT beams along the z direction (see figure \ref{fig:mot}) are nearly superimposed with the slowing-laser beam and merged with the atomic beam. The $\sigma_+$ MOT beam is directed by the mirror M at $45^{\circ}$ incidence and the glass viewport placed on the vacuum chamber 1 (see figure \ref{fig:expsetup}). It propagates along the z-axis through the vacuum chamber and the Zeeman slowers. The two contrapropagating beams along the z-axis are focused onto the mirror M with an edge separated by 1 cm from the center of the atomic beam. In this geometry, both the $\sigma_+$ and the $\sigma_-$ of the MOT beams along the atomic beam axis affect the slowing process. Consequently, several precautions such as the use of a repumper beam are required for optimization of the MOT as explained in the following section.

\subsection{Optimization of the MOT}
\begin{figure}[htb]
\begin{center}
\includegraphics[height=4.7cm,width=9cm]{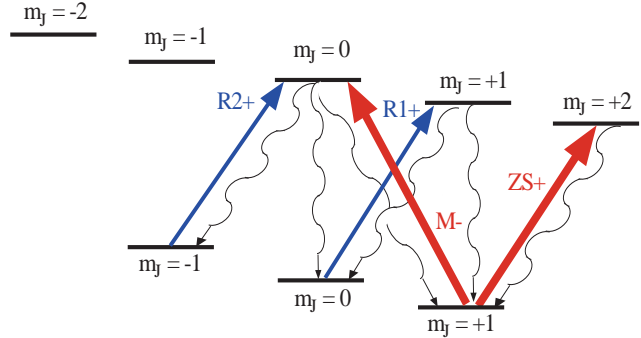}
\end{center}
\caption{\footnotesize Repumping scheme between $2^{3}S_1$ and $2^{3}P_2$ states of helium. ZS+ corresponds to the Zeeman slower beam,$\sigma^+$ polarized, M- to the $\sigma^-$ MOT beam along the atomic beam, and R1+ and R2+ are the two repumping transitions required to bring the atoms back into the $m_J=+1$ sublevel and restore the slowing process.}
\label{fig:repump}
\end{figure}

On one hand, the $\sigma_+$ MOT beam along the z-axis is resonant with the atoms at a given position in the slowing magnet. It can thus be absorbed by the atoms and accelerate them. On the other hand, the $\sigma_-$ MOT beam along the z-axis is likely to depolarize the traveling atoms at another position. These two effects have to be corrected for. The slowing beam detuning from resonance is $\delta_{slo}/2\pi=-240$ MHz from resonance (beam B2 in fig. \ref{fig:lassetup}). During the slowing process, the velocity of the atom decreases according to the following equation (\ref{eq:vslo})
\begin{equation}
\delta_{slo}+kv=\mu_b B/\hbar 
\label{eq:vslo} 
\end{equation}
where $B$ and $v$ are the projection along the z-axis of the magnetic field and atom velocity respectively. The atoms are spin polarized in the $m_J=+1$ level during the slowing process, cycling between the states $2^{3}S_1,g_s=2, m_J=+1$ and $2^{3}P_2,g_p=3/2, m_J=+2$ (see fig. \ref{fig:repump}).\\
The $\sigma_+$ MOT beam, parallel to the atomic beam (see fig. \ref{fig:mot}), can also induce transitions between these magnetic sublevels, if the following resonance condition is fulfilled:
\begin{equation}
\delta_{mot}-kv=\mu_b B/\hbar  
\label{eq:vmot} 
\end{equation}
Eq. (\ref{eq:vslo}) and (\ref{eq:vmot}) are both satisfied for 
\begin{equation}
2kv_{+}=\delta_{mot} -\delta_{slo} 
\label{eq:v+} 
\end{equation}
which gives $v_{+}=105$ m/s. So, when the velocity becomes $v_{+}$, which occurs before the end of the slowing process, in the second part of the Zeeman slower, the $\sigma_+$ MOT beam accelerates the atoms. The net effect results from the intensity unbalance between the $\sigma_+$ MOT beam and the slowing beam. If the intensity of the slowing beam is less than the MOT beam intensity, the $\sigma_+$ MOT beam accelerates the atoms so much that the slowing process is stopped. One needs to adjust the intensity of the slowing beam to be higher than the intensity of the MOT beam to prevent this undesirable phenomenon. We typically use 15 mW/cm$^2$ for the slowing beam, which corresponds to 1.5 times the intensity of each of the MOT beams.

In addition, the $\sigma_-$ MOT beam along the z-axis can induce transitions between $2^{3}S_1, m_J=+1$ and $2^{3}P_2, m_J=0$ sublevels, which depolarize the atoms when they decay to the $2^{3}S_1, m_J=0$ and $m_J=-1$ sublevels (see fig. \ref{fig:repump}). Once the atoms have decayed, they are no longer resonant with the slowing beam and the slowing process is stopped. This actually happens when the following resonance condition is fulfilled:
\begin{equation}
\delta_{mot}+kv=-2\mu_b B/\hbar  
\label{eq:vmot2} 
\end{equation}
Eq. (\ref{eq:vslo}) and (\ref{eq:vmot2}) are both satisfied for 
\begin{equation}
3kv_{-}=-(\delta_{mot}+2\delta_{slo}) 
\label{eq:v+} 
\end{equation}
which gives $v_{-}=190$ m/s. This velocity is also reached in the second part of the Zeeman slower.
To avoid this problem, we repump the atoms from the $m_J=0$ and $m_J=-1$ sublevels back to the $m_J=+1$ sublevel. Two repumping beams are both $\sigma_+$ polarized and resonant with the transitions $2^{3}S_1, m_J=-1\rightarrow2^{3}P_2, m_J=0$ and $2^{3}S_1, m_J=0\rightarrow2^{3}P_2, m_J=+1$, at exactly the same magnetic field and the same atomic velocity at which the depumping happens (see fig. \ref{fig:repump}). One calculates that the required frequencies are detuned by -272.5 MHz and -305 MHz from the atomic resonance at zero magnetic field. To generate the required frequencies, we lock a unique additional DBR laser tuned -289 MHz from resonance. We RF-modulate the diode current at 16 MHz. This generates sidebands into its spectrum. The level of modulation is optimized to get the maximum intensity into the first two lateral sidebands, whose frequencies are the ones required for repumping. The power in the repumping beam is approximately 20 mW. It is checked using the absorption measurement explained earlier that the repumping process brings back nearly the same flux of slow atoms as in the abscence of the MOT beams.

\subsection{Characterization of the MOT}
The atoms are finally confined in a magneto-optical trap. Two cylindrical coils, separated by 5.2 cm along the y-axis (see figure \ref{fig:mot}), create a magnetic gradient of 40 G/cm along the symmetry axis for a given current of 5A. The repumper beam typically increases the number of trapped atoms by a factor of 3. Losses are dominated by intra MOT Penning collisions \cite{Tol99,Kuma99,Brow00,Per00}. In this regime, the number of trapped atoms goes as the square root of the loading rate. This increase in the number of atoms implies that the loading rate is increased by a factor of $3^2=9$. Using the repumper beam, we routinely trap approximately $10^9$ atoms, in a volume of 0.1 cm$^3$, at a temperature of approximately 1 mK. The temperature is measured by a time of flight technique \cite{Per00}. The number of trapped atoms is inferred from the measurement of the absorption in a 1 cm diameter probe laser beam, intense enough to saturate the transition. The size of the MOT is measured by absorption imaging on a CCD Camera.

\section{Conclusion}
\label{sec:concl}
In this article, we report on a new experiment aiming at reaching BEC with metastable helium atoms. We demonstrate the efficiency and the robustness of a new discharge source of metastable atoms and of a bright laser setup using a high power fiber amplifier at 1.083 nm for the manipulation of atoms. For the MOT the original trapping scheme requires an additional laser beam used to repump the atoms during the slowing process. The atomic cloud is trapped at the center of a small quartz cell. The present setup has several advantages. First, it gives a good optical access to the atomic cloud. This allows to further trap atoms in a strongly confining magnetic trap placed as close possible to the cell as. Secondly, it allows to reach extremely low pressures inside the small volume of the cell. However, the present setup makes it difficult to use ion detectors or channel plates to detect the metastable atoms. Further developements of the experiment could include such detectors in an appropriate cell. Detection is performed in the present setup by purely optical means. The infrared line or other visible lines for which CCD cameras have a better efficiency can be use for detection. The present experiment allows to routinely trap approximately $10^9$ helium atoms in the $2^3S_1$ metastable state inside a MOT of 2 mm rms radius.\\
The setup is currently being modified to add magnetic coils for a magnetostatic trap that will be used in the search for BEC. This Ioffe type trap consists of three asymmetric coils plus two large compensation Helmholtz coils, giving a field configuration similar to the QUIC trap \cite{Essl98}. The gradients are approximately 280 G/cm, the curvature is 200 G/cm$^2$ and the depth of 33 mK for a current of 50 A. Before loading in the QUIC trap, the atoms will be first cooled into a molasse phase, where the field gradient of the MOT is turned off, which should allow to reach lower temperatures (50 to 100 $\mu$K). When a large density of ultracold atoms is loaded into the Ioffe trap, it will be possible to check the theoretical predictions of \cite{Shlyap96,Venturi99} on elastic and inelastic collision rates between metastable atoms at very low temperature. The measured collision rates will then indicate whether one can achieve BEC using evaporative cooling, as successfully used with alkali atoms.\\

\noindent {\bf Acknowledgments:}
The authors thank C. Cohen-Tannoudji for very helpful discussions, and for his input in the experiment.\\ 

$^a$ Permanent address: Laboratoire de Physique des Lasers, UMR 7538 du CNRS, Universit\'e Paris Nord, Avenue J.B. Cl\'ement, 93430 Villetaneuse, France.

$^b$ Permanent address: Institute of Opto-Electronics, Shanxi University, 36 Wucheng Road, Taiyuan, Shanxi 030006, China.

$^c$ Permanent address : Dept. of Physics, Univ. of Perugia, Via Pascoli, Perugia, Italy; Lens and INFM, L.go E. Fermi 2, Firenze, Italy

$^d$ Present address : Universit$\ddot{\rm{a}}$t Hannover, Welfengarten 1, D-30167 Hannover, Germany.

$^e$ Permanent address : TIFR, Homi Bhabha Road, Mumbai 400005, India.

$^*$ Unit\'e de Recherche de l'Ecole Normale Sup\'erieure et de
l'Universit\'e Pierre et Marie Curie, associ\'ee au CNRS (UMR 8552).

\end{document}